\documentclass[preprint,review,12pt,sort&compress]{elsarticle}

\usepackage{amsfonts, amsmath, amssymb, amsthm}
\usepackage{amssymb}
\usepackage{nomencl}
\usepackage{calc}
\usepackage{empheq}
\usepackage{color}
\usepackage{transparent}  
\usepackage{mathtools}
\usepackage{empheq}
\usepackage{bm}
\usepackage{lineno}
\usepackage{array}
\usepackage{multirow}
\newcolumntype{P}[1]{>{\centering\arraybackslash}p{#1}}   
\newcolumntype{M}[1]{>{\centering\arraybackslash}m{#1}}   
\newcolumntype{L}[1]{>{\raggedright\let\newline\\\arraybackslash\hspace{0pt}}m{#1}} 
\newcolumntype{C}[1]{>{\centering\let\newline\\\arraybackslash\hspace{0pt}}m{#1}}
\newcolumntype{R}[1]{>{\raggedleft\let\newline\\\arraybackslash\hspace{0pt}}m{#1}}
\usepackage{graphicx}
\usepackage{caption}
\usepackage{subfig}
\usepackage{citesort}

\DeclareMathOperator*{\argmin}{arg\:min}

\begin{document}

\begin{frontmatter}



\title{A Multiscale Diffuse-Interface Model for Two-Phase Flow in Porous Media}


\author{M. Shokrpour Roudbari\corref{cor1}, E.H. van Brummelen and C.V. Verhoosel}
\address{Eindhoven University of Technology -- Multiscale Engineering Fluid Dynamics chair, 
P.O. Box 513, 5600 MB -- Eindhoven, The Netherlands}
\ead{m.shokrpour.roudbari@tue.nl}

\begin{abstract}
In this paper we consider a multiscale phase-field model for capillarity-driven flows in
porous media. The presented model constitutes a reduction of the conventional 
Navier-Stokes-Cahn-Hilliard phase-field model, valid in situations where interest 
is restricted to dynamical and equilibrium behavior in an aggregated sense, rather 
than a precise description of microscale flow phenomena. The
model is based on averaging of the equation of motion, thereby yielding
a significant reduction in the complexity of the underlying Navier-Stokes-Cahn-Hilliard equations,
while retaining its macroscopic dynamical and equilibrium properties. Numerical results 
are presented for the representative 
2\nobreakdash-dimensional capillary-rise problem pertaining to two closely spaced vertical plates with 
both identical and disparate wetting properties. Comparison with analytical solutions for these test cases corroborates the accuracy of the presented multiscale model. In addition, we present results for 
a capillary-rise problem with a non-trivial geometry corresponding to a porous medium.
\end{abstract}

\end{frontmatter}

\newcommand{\U}{\ensuremath{\bm{u}}}  
\newcommand{\Uw }{\ensuremath{\bm{U}_w}}
\newcommand{\Uc }{\ensuremath{\bm{U}_c}}
\newcommand{\X }{\ensuremath{\bm{x}}}
\newcommand{\N }{\ensuremath{\bm{n}}}
\newcommand{\D }{\ensuremath{\bm{D}}}
\newcommand{\J }{\ensuremath{\bm{j}}}
\newcommand{\HH }{\ensuremath{\bm{h}}}
\newcommand{\TT }{\ensuremath{\mathcal{T}}}
\newcommand{\F }{\ensuremath{\bm{f}}}
\newcommand{\KK }{\ensuremath{\bm{K}}}
\newcommand{\G }{\ensuremath{\bm{g}}}
\newcommand{\II }{\ensuremath{\bm{I}}}
\newcommand{\CC }{\ensuremath{\bm{C}}}
\newcommand{\Top }{\mathrm{on} \; \ensuremath{\Gamma}_\mathrm{top}}
\newcommand{\Bot }{\mathrm{on} \; \ensuremath{\Gamma}_\mathrm{bottom}}
\newcommand{\Wall }{\mathrm{on} \; \ensuremath{\Gamma}_\mathrm{wall}}
\newcommand{\TopBot }{\mathrm{on} \; \ensuremath{\Gamma}_\mathrm{top} \cup \ensuremath{\Gamma}_\mathrm{bottom}}
\newcommand{\IR}{\mathbb{R}}
\let\oldomega\omega
\renewcommand{\omega}{\bm{\oldomega}}
\let\Oldnabla\nabla
\renewcommand{\nabla}{\bm{\Oldnabla}}
\let\oldtau\tau
\renewcommand{\tau}{\bm{\oldtau}}
\let\Oldvarsigma\varsigma
\renewcommand{\varsigma}{\bm{\Oldvarsigma}}
\let\Oldzeta\zeta
\renewcommand{\zeta}{\bm{\Oldzeta}}
\let\Oldalpha\alpha
\renewcommand{\alpha}{\bm{\Oldalpha}}
\newcommand{\SEC}[1]{\ref{sec:#1}}
\newcommand{\TAB}[1]{\ref{tab:#1}}
\newcommand{\FIG}[1]{\ref{fig:#1}}
\newcommand{\EQ}[1]{(\ref{eq:#1})}
\newcommand{\meas}{\mathrm{meas}}
\newcommand{\dd}{\mathrm{d}}
\renewcommand{\div}{\mathrm{div}{}}
\newcommand{\Fmix}{\mathcal{F}_{\textsc{m}}}
\newcommand{\Fwall}{\mathcal{F}_{\textsc{w}}}
\newcommand{\fmix}{f_{\textsc{m}}}
\newcommand{\Fg}{\mathcal{F}_{\textsc{g}}}
\newcommand{\Ftot}{\mathcal{F}}
\newcommand{\Kkin}{\mathcal{K}}
\newcommand{\mean}[1]{\langle{#1}\rangle}
\newcommand{\ave}[1]{\langle{#1}\rangle^{\star}}
\newcommand{\Pe}{\mathrm{Pe}}
\newcommand{\Cn}{\mathrm{Cn}}
\newcommand{\Bo}{\mathrm{Bo}}
\newcommand{\BoSharp}{\mathrm{Bo}^{\star}}
\newcommand{\rhol}{\rho_{\textsc{l}}}
\newcommand{\rhoa}{\rho_{\textsc{a}}}
\newcommand{\thetae}{\theta_{\textsc{e}}}
\newcommand{\sigmasa}{\sigma_{\textsc{sa}}}
\newcommand{\sigmasl}{\sigma_{\textsc{sl}}}
\newcommand{\sigmasf}{\sigma_{\textsc{sf}}}
\newcommand{\Sigmasf}{\Sigma_{\textsc{sf}}}
\newcommand{\he}{h_{\textsc{e}}}
\newcommand{\phie}{\phi_{\textsc{e}}}
\newcommand{\mua}{\mu_{\textsc{A}}}
\newcommand{\mul}{\mu_{\textsc{L}}}
\newcommand{\Gammaw}{\Gamma_{\textsc{w}}}
\newcommand{\Gammaio}{\Gamma_{\textsc{io}}}
\newcommand{\Tf}{t_{\textsc{f}}}
\newcommand{\Psic}{\Psi_{\textsc{c}}}
\newcommand{\Psie}{\Psi_{\textsc{e}}}
\newcommand{\Nw}{\N_{\textsc{w}}}

\section{Introduction}
\label{sec:Intro}
Capillarity-driven flows in porous media form a common occurrence in daily life
and in many applications in science and engineering. Despite significant
progress in recent years, modeling and simulation of such two-phase flows 
with moving interfaces and moving contact lines in complex geometries remains 
a challenging problem. Reliable numerical simulations are indispensable in the 
investigation of many industrial and natural phenomena related to porous-media flows, which 
generally have an inherently complex nature and demand encyclopedic experiments. 
For problems for which experimental investigations are expensive or hazardous 
(e.g. in nuclear power stations), problems with very large scales (e.g. geological 
and geophysical problems) or conversely problems with very small scales (e.g. 
in micro systems) or problems which exhibit a high-dimensional parameter space, 
experimental investigation is generally not feasible and the development of 
rigorous models is necessitated; see~\cite{vafai2005handbook,coutelieris2012transport} and the references therein.

Various modeling paradigms exist for porous-media flows, which can be classified
in different categories. Categorization can proceed for instance according to the 
ratio of the characteristic length scale of observation over the pore scale, as microscopic or macroscopic, or according to the number of fluid constituents as single- or multi-component. 
More specifically, two-phase porous-media flows in which one of the constituents 
consists of a gas phase and the other of a liquid phase, are typically classified as unsaturated.
Further subclassifications are possible, for instance for multi-component porous-media flows according to the viscosity ratio of the fluid components. A detailed review of these different model classes and their characteristics is provided in~\citep{vafai2005handbook}. The most common macroscopic model in porous-media flows is the celebrated Darcy model; see~\cite{bear2013dynamics,dybbs1984new} for comprehensive overviews. The Darcy model was originally conceived empirically~\cite{Darcy:1856yu} as a model for fully saturated porous-media flows, and was later re-established via homogenization 
procedures~\cite{neuman1977theoretical,whitaker1986flow}. Generalization of the Darcy model to a macroscopic model for porous-media flows in the unsaturated regime with a realistic representation of capillary effects remains an outstanding challenge, on account of the complexity of the wetting phenomena that occur~\cite{papatzacos2002macroscopic} and the intricacies of the tri-phasic configurations; see for instance~\cite{Szymkiewicz:2013jk}.

It appears that modeling and simulation of two-phase porous-media flows on a microscopic scale 
has received only scant consideration so far. Diffuse-interface (or {\em phase-field\/}) 
models~\cite{emmerich2003diffuse,anderson1998diffuse,lowengrub1998quasi} in principle provide a
cogent microscale modeling paradigm for two-phase flows in porous media. Diffuse-interface models
have emerged over the past years as a class of comprehensive and versatile models for
multi-component flows, enabling realistic descriptions of complicated physical phenomena such as
evaporation and condensation~\cite{Korteweg:1901kn,Liu:2013vn}, topological changes of fluid-fluid interfaces due to coalescence and fissuring~\cite{Jacqmin1999calculation,Yue:2006comp,lowengrub1998quasi}, contact-line motion at fluid-solid interfaces~\cite{seppecher1996moving,jacqmin2000contact,yue2011wall}, and 
elasto-capillary effects of complex fluids on elastic solid substrates~\cite{Brummelen:2015qv}. 
Diffuse-interface models intrinsically bypass the contact-line paradox of sharp-interface
models~\cite{huh1971hydrodynamic,jacqmin2004onset,seppecher1996moving}, thereby enabling
the description of preferential-wetting phenomena. Another compelling property of diffuse-interface models in the context of numerical simulation of  two-phase flows in porous media is the intrinsic description of the fluid-fluid interface, which significantly facilitates the treatment of complex geometries.
  
Diffuse-interface models for binary-fluid flows generally combine the Cahn--Hilliard equations~\cite{Cahn:1958fk}, which yield a model for phase-separation in binary mixtures, with the Navier--Stokes equations. 
The Navier--Stokes--Cahn--Hilliard (NSCH) system provides a complete model for binary-fluid flows and, in combination with appropriate wetting boundary conditions~\cite{jacqmin2000contact,yue2011wall}, 
gives a detailed description of flow and wetting phenomena. Numerical simulation of the NSCH equations represents many challenges, however. 
The equations are highly nonlinear and exhibit an inherent multiscale character, both in space and in time. If the NSCH equations serve as a model for capillarity-driven porous-media flows,
then these complications are further compounded by the geometric complexity of porous media.
In many cases of practical interest, however, one is not concerned with the fine details
 of the flow field and the meniscus evolution, and a reduced model for the flow suffices, 
 provided that an adequate macroscopic description of the meniscus evolution is obtained, and 
 that the equilibrium properties of the NSCH equations are retained at a macroscopic level.
The objective of this paper is to develop such a model. Let us note that as opposed to the recent phase-field models for unsaturated porous-media flow in~\cite{Cueto-Felgueroso:2009ud,Cueto-Felgueroso:2009dn}, 
the model presented in this work aims to provide a microscopic description of the
meniscus propagation and the flow that explicitly accounts for the geometry of the porous medium.

The remainder of this paper is arranged as follows. Section~\SEC{ProbStat} provides a
description of the considered problem.
Section~\SEC{DiffInt} surveys the main aspects of diffuse-interface models for binary-fluid flows with wetting boundary conditions.
In Section~\SEC{MultiScale} we develop the multiscale diffuse-interface binary-fluid flow model for
capillarity-driven flows. Section~\SEC{NumExp} presents numerical experiments on the basis of the
multiscale model. Section~\SEC{ConcRem} presents a retrospection and conclusions.
  %
\section{Problem statement} 
\label{sec:ProbStat}
To provide a context for the models considered in this paper, in this section we 
first establish a multiscale perspective on capillary flow in porous media. 
Next, we regard the essential features of capillary flows on the basis of the 
the generic problem of capillary action of a fluid in a narrow tube.  

\subsection{A multiscale perspective on capillary flow in porous media}
\begin{figure}
\begin{center}
\includegraphics[width=\textwidth]{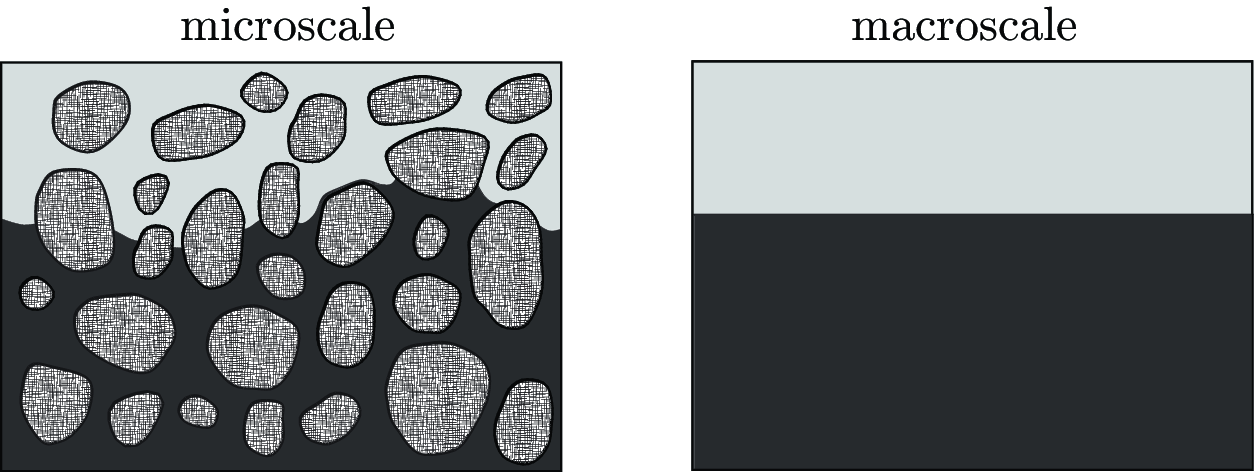}
\caption{Illustration of capillary flow in a porous medium on the microscale ({\em left\/}) and
its macroscale representation ({\em right\/}).}
\label{fig:Sketch}
\end{center}
\end{figure}

We consider the flow of a binary-fluid confined by a rigid porous solid; cf. 
Figure~\FIG{Sketch} ({\em left\/}). The solid exhibits
preferential-wetting properties for the two fluid components, i.e. the fluid-solid
surface tension is distinct for the two fluid components. In addition to the 
surface-tension effects at the fluid-fluid interface and at the fluid-solid interface, 
we assume that the fluid 
motion is induced by a potential field, for instance a gravity field
or a pressure gradient. 

The preferential-wetting property of the solid in combination with the geometric
complexity of the pore volume generally leads to a profoundly complicated evolution of the fluid-fluid meniscus on the microscale associated with the pore size. Moreover, on the microscale the meniscus can display many stable or meta-stable configurations and, accordingly, 
microscopic equilibrium configurations are generally non-unique. 

A macroscale representation of the considered binary-fluid flow is obtained by appropriately averaging 
the microscopic phase distribution of the two fluid components on a length scale that is sufficiently 
large compared to the pore size; see Figure~\FIG{Sketch} ({\em right\/}). The complicated
evolution of the meniscus on the microscale and its inherent sensitivity to perturbations do not
generally transfer to the macroscale, by virtue of the fact that local variations of the 
microscopic phase distribution 
are averaged out in the corresponding macroscopic distribution. 

The insensitivity of the macroscopic phase distribution to local variations in the underlying
microscopic phase distribution, implies that two models that exhibit distinct behavior on the
microscale, can still yield essentially identical macroscale behavior. Accordingly, one can
envisage a simplified binary-fluid flow model based on an aggregated representation of the 
flow dynamics, that still provides an accurate description of the meniscus evolution on
the macroscale, and that retains the macroscopic equilibrium properties of the binary-fluid flow.

This paper is concerned with the derivation of a multiscale diffuse-interface binary-fluid 
flow model for capillarity-driven flow in porous media. The model is based on a reduction of the
Navier-Stokes-Cahn-Hilliard (NSCH) phase-field model for binary-fluid flows, subject to preferential-wetting boundary conditions. The reduction essentially comprises an aggregation of momentum balance.
The reduced model is significantly less complex than the underlying NSCH equations, and retains 
the dynamical and equilibrium behavior of the NSCH model at the macroscale.

\subsection{Model problem: capillary flow in a cylinder}
\label{sec:CapTube}
\begin{figure}
\centering
\includegraphics[width=0.375\textwidth]{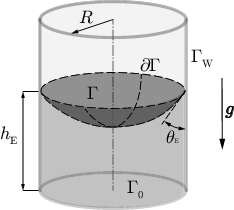}
\caption{Illustration of capillary rise in a cylindrical tube.}
\label{fig:CapTubeSketch}
\end{figure}

To provide a context for the models considered in this paper, we first
consider the generic problem of capillary action of a fluid in a narrow tube.  
For a detailed treatment of this phenomenon, we refer to the celebrated work
of De Gennes~\cite{de2013capillarity}. We regard the liquid and ambient fluids 
as two components of a binary complex fluid. Subscripts $(\cdot)_{\textsc{l}}$ and $(\cdot)_{\textsc{a}}$ will be used as identifiers for the liquid and ambient fluid (gas), respectively. A subscript $(\cdot)_{\textsc{s}}$ is used as an identifier for the solid tube. If the meniscus (fluid-fluid interface) is in equilibrium and in contact with the solid surface, then an equilibrium contact angle occurs in accordance with Young's equation:
\begin{equation}
\label{eq:CA}
\cos(\thetae) = \frac{\sigmasa- \sigmasl}{\sigma}  
\end{equation}
where $\sigmasl,\sigmasa$ and $\sigma$ refer to the liquid-solid, ambient-solid 
and liquid-ambient surface tensions, respectively. The contact angle between the meniscus and the fluid-solid interface is interior to the liquid. 
If $\thetae < 90^{\circ}$ the solid surface has energetic-affinity to contact with the liquid component relative to the ambient fluid. The meniscus then admits 
a convex shape (cf. Figure~\FIG{CapTubeSketch}) and propagation of the fluid in the tube leads to wetting of a larger surface area of the solid and a corresponding reduction of energy. In the presence of gravity and under the general assumption that the liquid is relatively heavy compared to the ambient gas, the reduction in surface energy can surpass gravity, and the liquid can rise up the tube.
%

The aforementioned capillary rise can alternatively be explained via static equilibrium of the
meniscus and the liquid column. On account of $\sigmasa>\sigmasl$, the meniscus
experiences a line-tension of $\sigmasa-\sigmasl$, normal to the contact line and 
tangential to the solid surface. This line tension is equilibrated by the pressure difference 
at the meniscus in accordance with the Young-Laplace relation,
which stipulates that the pressure in the ambient liquid above the meniscus exceeds the 
pressure in the liquid just below the meniscus by $\Delta{}p=\sigma{}\varkappa$ with~$\varkappa$ the additive curvature of the surface~\cite{de2013capillarity}. Hence, it must hold that
$(\sigmasa-\sigmasl)\,\meas(\partial\Gamma)=\sigma\varkappa\,\meas(\Gamma_0)$ with
$\Gamma$ the meniscus, $\partial\Gamma$ its boundary, and $\Gamma_0$ the projection
of~$\Gamma$ onto a plane normal to the tube; cf. Figure~\FIG{CapTubeSketch}. Static equilibrium of the liquid column in turn requires that the pressure at the base of the column balances the pressure 
just below the meniscus and the excess gravitational force on the liquid column relative to the
gravitational force of the ambient fluid. Hence, we have the sequence of identities 
\begin{equation}
\label{eq:h_e1}
(\rhol-\rhoa)\,\he{}g\,\meas(\Gamma_0)=\sigma\varkappa\,\meas(\Gamma_0)=(\sigmasa-\sigmasl)\,\meas(\partial\Gamma)
\end{equation}
with $\he$ as the equilibrium height and $g$ as gravitational acceleration. The equilibrium
height can be extracted from~\EQ{CA} and~\EQ{h_e1} as 
\begin{equation}
\label{eq:h_e}
\he=\frac{\sigma\,\cos(\thetae)\,\meas(\partial\Gamma)}{(\rhol-\rhoa)g\,\meas(\Gamma_0)}
\end{equation}
One can infer that both for a circular tube with radius~$R$ (in 3D) and
for 2 parallel plates separated by a distance $R$ (in 2D), 
the ratio of the circumference $\meas(\partial\Gamma)$ to the projected 
area $\meas(\Gamma_0)$ is~$2/R$.

\begin{figure}
\begin{center}
\includegraphics[width=0.6\textwidth]{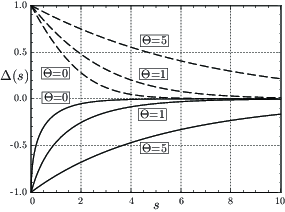}
\caption{Evolution of the meniscus height according to~\EQ{Delta} for $(\pm,c)=(+,e), (- -)$ and $(\pm,c)=(-,1), (\mbox{---}) $.}
\label{fig:Delta}
\end{center}
\end{figure}

The above consideration of the equilibrium configuration of the meniscus based on contact-line and gravitational forces can be extended to the dynamic behavior of the meniscus. We assume
that the viscous dissipation in the binary fluid is proportional to its (longitudinal) velocity
and to the average viscosity of the binary fluid. Denoting the viscosity of the liquid and 
ambient fluid by~$\mul$ and~$\mua$, respectively, viscous dissipation then induces a
force
\begin{equation}
F_{\textsc{D}}=-c_{\textsc{D}}\big(\mul{}h+\mua(H-h)\big)d_th
\end{equation}
with $h:=h(t)$ as the time-dependent height of the meniscus, $H$ the length of the capillary tube,
and~$c_{\textsc{D}}>0$ a constant.
Ignoring inertial effects, equilibrium of forces leads to the following ordinary differential equation
for the meniscus height:
\begin{equation}
\label{eq:Dynamic1}
(\sigmasa-\sigmasl)\meas(\partial\Gamma)
-(\rhol{}-\rhoa)h\,g\,\meas(\Gamma_0)
-c_{\mathrm{D}}\big((\mul-\mua){}h+\mua{}H\big)d_th=0
\end{equation}
Invoking relation~\EQ{h_e1} between the equilibrium height and the capillary and gravitational forces,
and introducing 
\begin{equation}
\Lambda=
\frac{c_{\textsc{D}}(\mul-\mua)}{(\rhol{}-\rhoa)g\,\meas(\Gamma_0)}
\qquad
\Theta=\frac{H\mua}{\he(\mul-\mua)}
\qquad
\Delta(s)=\frac{h(\Lambda{}s)}{\he}-1
\end{equation}
we can condense~\EQ{Dynamic1} into:
\begin{equation}
\label{eq:DeltaODE}
\Delta+\big(\Delta+1+\Theta\big)\Delta'=0
\end{equation}
The ordinary-differential equation~\EQ{DeltaODE} can be solved to obtain the general solution:
\begin{equation}
\label{eq:Delta}
\Delta(s)=(1+\Theta)\,W\bigg(\pm\frac{1}{1+\Theta}\exp\Big(\frac{c-s}{1+\Theta}\Big)\bigg)
\end{equation}
with $W:[-1/e,\infty)\to[-1,\infty)$ the Lambert W-function (see~\cite{Corless:1996kl}) and $c$ an arbitrary constant. 
The $\pm$ symbol represents two distinct solutions, viz. a strictly positive ($+$) and a 
strictly negative ($-$) solution. It is to be noted that~$c$ merely effects a translation 
of the solution.
To elucidate the evolution of the meniscus height according to~\EQ{Delta},
Figure~\FIG{Delta} plots integral curves of~\EQ{Delta} for $\Theta\in\{0,1,5\}$
and for $\pm=-$ and $c=-1$ and for $\pm=+$ and $c=e$. The values of $c$ have been selected
such that $\Delta(0)=\pm{}1$. On can observe that $\Delta(s)$ tends to~$0$ as~$s\to\infty$,
in accordance with the approach to equilibrium of the meniscus. Moreover, for larger values
of~$\Theta$, corresponding to a more pronounced effect of ambient viscosity, the approach
to equilibrium is retarded.
%
\\

\section{Diffuse-interface models for capillary flow}
\label{sec:DiffInt}
In Sections~\SEC{Bulk}--\SEC{CHE} we review the standard diffuse-interface model for surface tension and wetting. Section~\SEC{NSCH} presents
the Navier-Stokes-Cahn-Hilliard model for binary-fluid flows, which serves as a comprehensive background model for the derivation of our reduced multiscale model in Section~\SEC{MultiScale}.

\subsection{Bulk free energy of a binary fluid with a diffuse interface}
\label{sec:Bulk}
The above consideration of the capillary-tube setup conveys that any meaningful
model of capillary flow must be able to account for surface tension and contact-line tension.
We consider here a diffuse-interface representation of these effects. We first regard the
bulk free energy of a binary fluid.

Let $\Omega\subset\IR^d$ ($d=2,3$) denote a domain occupied by a binary-mixture of 
two immiscible fluids. Denoting by $\phi:\Omega\to\IR$ an order parameter, the 
mixture is modelled by the Ginzburg-Landau-type energy functional:
    %
    \begin{equation}
    \label{eq:Fmix}
      \Fmix(\phi) = \int_\Omega \fmix{(\phi,\nabla \phi)} \; \dd{V}      
    \end{equation}
    with $\dd{}V$ as volume measure. The free-energy density $\fmix$ in~\EQ{Fmix} is of the general form:
    %
    \begin{equation}
      \fmix(\phi, \nabla \phi) = \tfrac{1}{2}P_1|\nabla \phi|^2  + P_2 \Psi(\phi)
      \label{eq:fDens}
    \end{equation}
    The model parameters~$P_1$ and~$P_2$ depend on the choice of the potential
    $\Psi(\phi)$ and on the definition the 
    interface; see e.g.~\cite{jacqmin2000contact}. The terms
    $\tfrac{1}{2}P_1|\nabla \phi|^2$ and  $P_2 \Psi(\phi)$ in \EQ{fDens} correspond to the 
    free-energy density due to the phase gradient and the bulk free energy, respectively.
    In fact, the potential~$\Psi$ models the immiscibility of the two fluid components.
    It has two local minima at $\xi^{\pm}$, corresponding to the two stable phases of the 
    fluid~\cite{jacqmin2000contact}. A standard expression for the potential~$\Psi$ 
    is the double-well function:
    %
    \begin{equation}
    \label{eq:Psi}
      \Psi(\phi) = \tfrac{1}{4}a_1\big(\phi+\sqrt{a_2/a_1}\big)^2\big(\phi-\sqrt{a_2/a_1}\big)^2 
    \end{equation}
The minima of $\Psi$ that indicate the pure phases occur at $\xi^{\pm}=\pm\sqrt{a_2/a_1}$. Intermediate values of the phase variable, $\phi\in(\xi^-,\xi^+)$, indicate local
co-existence of the two phases. 

    According to the van-der-Waals hypothesis, the equilibrium interface profiles are 
    those which minimize the free energy. A standard variational argument then conveys that
    such equilibrium profiles satisfy the optimality condition:
\begin{equation}
\label{eq:G}
G:=-P_1\Oldnabla^2\phi+P_2\Psi'(\phi)=0
\end{equation}
    with $G$ as the chemical potential and $\Oldnabla^2$ the Laplace operator. 
    From~\EQ{G} it follows that for an essentially flat interface, the phase 
    variable in equilibrium assumes the following form:
    %
    \begin{equation}
      \phie(s) = \sqrt{\frac{a_2}{a_1}}\,\text{tanh}\bigg(\sqrt{\frac{a_2P_2}{2P_1}}\,s\bigg) 
      \label{eq:EqProfile}
    \end{equation}
    where $s$ represents an arc-length coordinate normal to the interface. 
    Noting that surface tension corresponds to the increase in the free energy 
    that accompanies an increase in surface area of the meniscus~\cite{de2013capillarity}, 
    the surface tension associated with the equilibrium profile~\EQ{EqProfile} of 
    the phase variable is
    %
    \begin{equation}
    \label{eq:Sigma}
    \begin{aligned}
      \sigma = \int_{-\infty}^{+\infty} \fmix (\phie, \nabla \phie) \,
      \mathrm{d}s
                 &= \int_{-\infty}^{+\infty} \tfrac{1}{2}P_1 (\phie'(s))^2
                  +P_2\Psi\big(\phie(s)\big)\,\mathrm{d}s
                  \\
                 &= \frac{2\sqrt{2}}{3}\frac{a_2}{a_1}\sqrt{a_2P_1P_2}
    \end{aligned}
    \end{equation}
    The expression for the equilibrium profile in~\EQ{EqProfile} conveys that there is no precise
    delineation of the interface thickness. It however also follows from~\EQ{EqProfile} that 
    the ratio $\ell=\sqrt{2P_1/(a_2P_2)}$ acts as a length scale for the thickness of the diffuse interface.
    For instance, according to~\EQ{EqProfile}, for any $\Oldalpha\in(0,1)$ the phase variable changes 
    from $\Oldalpha\xi^-$ to~$\Oldalpha\xi^+$ within a distance of $2\ell\,\mathrm{arctanh}(\Oldalpha)$.
    Without loss of generality, we select $a_1=a_2=1$, which implies that the pure phases are 
    identified by $\phi=\pm{}1$. If we moreover set $P_1=\tilde{\sigma}\epsilon$ 
    and~$P_2=\tilde{\sigma}/\epsilon$, then it follows from~\EQ{Sigma} that the surface tension
    induced by the diffuse interface is related to $\tilde{\sigma}$ by $3\,\sigma=2\sqrt{2}\,\tilde{\sigma}$ while $\epsilon$ is related to the length scale $\ell$ by $\ell=\sqrt{2}\,\epsilon$. The latter
    implies concretely that $90\%$ of the phase change occurs across a transition layer of approximately $4\epsilon$ thickness.
    
\subsection{Surface tension at fluid-solid interfaces}  
\label{sec:SurfaceTensionFS}  
    The interaction of the binary fluid with a contiguous solid wall $\Gammaw$ can be represented by
    an additional contribution to the free energy of the form
    \begin{equation}
    \label{eq:Fw}
    \Fwall(\phi)=
    \int_{\Gammaw}\sigmasf(\phi)\,\dd{S}
    \end{equation}
    where $\dd{S}$ denotes surface measure. The surface tension $\sigmasf$ of the binary fluid
    and the solid must interpolate between the solid-fluid surface tensions of the components. 
    In particular, identifying $\phi=+1$ (resp. $\phi=-1$) with the liquid (resp. ambient fluid), 
    it must hold 
    that $\sigmasf(+1)=\sigmasl$ and $\sigmasf(-1)=\sigmasa$. In additon, to ensure that all line tensions vanish in the pure fluid states, it must hold that $\lim_{s\to\pm{}1}\sigmasf'(s)=0$. The simplest relation in compliance with the aforementioned conditions is the cubic-spline interpolation:
    \begin{equation}
    \label{eq:sigmasf}
    \sigmasf(s)=\tfrac{1}{2}(\sigmasl+\sigmasa)+\tfrac{1}{4}(\sigmasa-\sigmasl)(s^3-3s)
    \end{equation}
    see also~\cite{yue2011wall,jacqmin2000contact}.
    
    Equilibrium of the binary-fluid in interaction with the solid wall is characterized by minimization of the total free energy, consisting of the bulk free energy~\EQ{Fmix} and the solid-wall 
    free energy~\EQ{Fw}. Under the assumption that the fluid-fluid interface is essentially flat, one can establish straightforwardly that in equilibrium the diffuse interface meets the solid wall at the equilibrium contact angle according to~\EQ{CA}. A standard variational argument conveys that in addition to~\EQ{G} it must hold that
     \begin{equation}
     \label{eq:wettingBC}
     \tilde{\sigma}\epsilon\,\partial_n\phi+\sigmasf'(\phi)=0\qquad\text{on }\Gammaw
     \end{equation}
     If the interface is essentially flat, then it exhibits the equilibrium profile~\EQ{EqProfile}. Denoting the unit normal vector to the fluid-solid interface exterior to the solid by $\Nw$ and
     the normal to the meniscus exterior to the liquid by $\N_{\Gamma}$, one can infer that $\partial_n\phi=(\Nw\cdot\N_{\Gamma})\phie'(s)$. Inverting the relation between~$\phi$ and~$s$ in~\EQ{EqProfile},
     we obtain the identities:
     \begin{multline}
     \epsilon\partial_n\phi=(\Nw\cdot\N_{\Gamma})\,\epsilon\phie'(s)=
     \frac{\Nw\cdot\N_{\Gamma}}{\sqrt{2}}\mathrm{sech}^2\bigg(\frac{s}{\sqrt{2}\epsilon}\bigg)
     \\
     =
     \frac{\Nw\cdot\N_{\Gamma}}{\sqrt{2}}\,\mathrm{sech}^2\big(\mathrm{arctanh}(\phi)\big)
     =\frac{\Nw\cdot\N_{\Gamma}}{\sqrt{2}}(1-\phi^2)
     \end{multline}
     where $\mathrm{sech}(\cdot)$ denotes the hyperbolic-secant function. From~\EQ{sigmasf} and~\EQ{wettingBC} it in turn follows that
     \begin{equation}
     \tilde{\sigma}\frac{\Nw\cdot\N_{\Gamma}}{\sqrt{2}}(1-\phi^2)-\frac{3(\sigmasa-\sigmasl)}{4}(1-\phi^2)=0
     \end{equation}
     and, hence,
     \begin{equation}
     \label{eq:FinalID1}
     \Nw\cdot\N_{\Gamma}=\frac{3(\sigmasa-\sigmasl)}{2\sqrt{2}\tilde{\sigma}}=
     \frac{\sigmasa-\sigmasl}{\sigma}=\cos(\theta_s)
     \end{equation}
     The penultimate identity follows from $3\,\sigma=2\sqrt{2}\,\tilde{\sigma}$. Noting that $\Nw\cdot\N_{\Gamma}$ corresponds to the cosine of the angle between the meniscus and the solid surface, interior to the liquid, Equation~\EQ{FinalID1} corroborates that in equilibrium the diffuse-interface binary fluid satisfies~\EQ{CA}.
     
\subsection{Cahn-Hilliard equation: dynamics and dissipation}
\label{sec:CHE}
A basic dynamical model for capillary flows can be derived by supplementing the total free energy with
a gradient-flow structure; see for instance~\cite{Taylor:1994uk}. The most common gradient-flow structure for the Ginzburg-Landau-type free energy~\EQ{Fmix} is the gradient flow with respect to the
$H^{-1}$-inner product, which leads to the Cahn-Hilliard equation:
\begin{equation}
\label{eq:CH}
\partial_t\phi= \gamma\Oldnabla^2{}G
\end{equation}
with $G$ the chemical potential~\EQ{G} and $\gamma$ a rate parameter referred to as {\em mobility\/}.
If~\EQ{CH} is provided with the following standard boundary conditions, in addition to~\EQ{wettingBC},
\begin{subequations}
\label{eq:CHBC}
\begin{align}
\partial_n\phi&=0\qquad\text{on }\partial\Omega\setminus\Gammaw
\\
\partial_nG&=0\qquad\text{on }\partial\Omega
\end{align}
\end{subequations}
then the solution to~\EQ{CH} subject to~\EQ{wettingBC} and~\EQ{CHBC} dissipates the total free energy,
i.e. $d_t(\Fmix(\phi)+\Fwall(\phi))\leq{}0$. In addition, the following equivalences hold:
\begin{equation}
\label{eq:equivalences0}
\begin{aligned}
d_t(\Fmix(\phi)+\Fwall(\phi))=0
&\quad\Leftrightarrow\quad
\Fmix(\phi)+\Fwall(\phi)\overset{!}{=}\mathrm{min}
\\
&\quad\Leftrightarrow\quad
\text{$\phi$ is an equilibrium solution}
\end{aligned}
\end{equation}
The ultimate condition in~\EQ{equivalences0} connotes that minimizers of the total free energy characterize meaningful equilibrium solutions, in the sense that such solutions display a well defined meniscus with a proper contact angle at the fluid-solid interface; see Sections~\SEC{Bulk}--\SEC{SurfaceTensionFS}.

Despite the fact that the Cahn-Hilliard equation~\EQ{CH} subject to~\EQ{wettingBC} and~\EQ{CHBC} displays the correct dissipation properties and equilibrium behavior, its merits as a dynamic capillary-flow model are limited: the equation does not provide an equation of motion based on conservation of momentum. Moreover,
if~$\tfrac{1}{2}(1+\phi)$ is given the usual interpretation of (liquid-)volume fraction, then~\EQ{CH}
cannot properly incorporate gravitational effects. With $\tfrac{1}{2}(1+\phi)$ as liquid-volume fraction,
the mixture density is
\begin{equation}
\label{eq:mixndens}
\rho(\phi)=\tfrac{1}{2}(1+\phi)\rhol+\tfrac{1}{2}(1-\phi)\rhoa
\end{equation}
and the corresponding gravitational energy of the binary fluid is
\begin{equation}
\label{eq:Fg}
\Fg=\int_{\Omega}\rho(\phi)(\G\cdot\X)\,\dd{V}
\end{equation}
with $\G=-g\J$ as gravitational acceleration. One easily infers that the variational derivative 
of~\EQ{Fg} is independent of~$\phi$ and linear in $\X$ and, therefore, the gravitational contribution
to the chemical potential vanishes under the Laplace operator in~\EQ{CH} and does not contribute
to the dynamics. The evolution of~$\phi$
under~\EQ{CH} and~\EQ{wettingBC} then dissipates the total free energy $\mathcal{F}=\Fmix+\Fwall+\Fg$,
but the equivalences in~\EQ{equivalences0} are violated and steady solutions of~\EQ{CH} do not
generally correspond to minizers of~$\Ftot$, nor to proper equilibrium solutions.

 
  \subsection{Navier-Stokes-Cahn-Hilliard equations}
  \label{sec:NSCH}
  To endow the diffuse-interface binary-fluid flow model with credible dynamic behavior, the Cahn-Hilliard equation~\EQ{CH} is supplemented with a transport velocity $\U$ that is itself subject to the Navier-Stokes equations. Reciprocally, the surface tension corresponding to the free energy~\EQ{Fmix} is 
  introduced as an additional contribution to the Navier-Stokes equation.  The resulting coupled equations are generally referred to as the Navier-Stokes-Cahn-Hilliard (NSCH) equations:
    \begin{subequations}
    \label{eq:NSCH}
    \begin{align}
      \partial_t\big(\rho(\phi)\U\big)+\div{}\big(\rho(\phi)\U\otimes\U+p\boldsymbol{I}+\zeta-\tau\big)
       &= \rho(\phi) \G \label{eq:NSCHmom}
       \\
      \div\U&=r(\phi)\label{eq:quasi}
      \\
      \partial_t \phi+ \div(\U\phi) &= \gamma \Oldnabla^2 G \\
      \label{eq:NSfull} 
      - \tilde{\sigma} \epsilon \Oldnabla^2 \phi + (\tilde{\sigma}/\epsilon)\Psi'(\phi)&=G
    \end{align}
    \end{subequations}
    with $\boldsymbol{I}$ the identity tensor, and $\tau = \mu(\phi)(\nabla \U + (\nabla \U)^T)$ and 
    $\zeta = \tilde{\sigma} \epsilon( \nabla\phi  \otimes  \nabla \phi)$
    the Cauchy stress and the capillary tensor, respectively.
    The right-hand side term in~\EQ{quasi} serves to represent different subtypes of
    the NSCH equations, e.g. quasi-incompressible models~\cite{lowengrub1998quasi} and 
   incompressible models~\cite{Abels:2012vn}. We will restrict 
    our considerations here to incompressible models and, accordingly, set~$r(\phi)=0$. 
      
    The NSCH equations~\EQ{NSCH} in conjunction with the wetting boundary condition~\EQ{wettingBC} provide a comprehensive model for binary-fluid flows, including fluid-fluid surface tension, capillary effects, gravitational effects, viscous dissipation, etc.; see Section~\SEC{Intro}.
    For proper subtypes of~\EQ{NSCH} it can moreover be established that the evolution of $\phi,\U$ subject to~\EQ{wettingBC} (and appropriate auxiliary boundary conditions) dissipates the convex functional 
    $\Ftot(\phi)+\Kkin(\phi,\U)$ 
    where $\Kkin(\phi,\U)=\int_{\Omega}\tfrac{1}{2}\rho(\phi)|\U|^2$ denotes the kinetic energy.
    In addition, the following equivalences hold:
\begin{equation}
\label{eq:equivalences1}
\begin{aligned}
d_t(\Ftot(\phi)+\Kkin(\phi,\U))=0
&\quad\Leftrightarrow\quad
\Ftot(\phi)+\Kkin(\phi,\U)\overset{!}{=}\mathrm{min}
\\
&\quad\Leftrightarrow\quad
\text{$(\phi,\U)$ is an equilibrium solution}
\end{aligned}
\end{equation}
If a standard no-slip boundary condition holds at~$\Gammaw$, then $\U$ vanishes in equilibrium. The
kinetic energy $\Kkin(\phi,\U)$ then vanishes as well and the second condition in~\EQ{equivalences1} conveys that the equilibrium order parameter minimizes the aggregated free energy~$\Ftot$. Stationary solutions to~\EQ{NSCH} in conjunction with~\EQ{wettingBC} (and appropriate auxiliary boundary conditions)
therefore display the contact angle~\EQ{CA}, the equilibrium interface profile in~\EQ{EqProfile} 
and the correct meniscus elevation.

    Numerical simulation of the NSCH system~\EQ{NSCH} represents many challenges. The equations are highly nonlinear and display an inherent multiscale character, both in space and in time. In Section~\SEC{MultiScale}, we develop a multiscale binary-fluid flow model that provides a coarse description of flow dynamics, but retains the equilibrium properties encoded by~\EQ{equivalences1}.

\section{A multiscale diffuse-interface binary-fluid flow model} 
\label{sec:MultiScale}
Our derivation of the multiscale diffuse-interface binary-fluid flow model is based on
a Slattery-Whitaker averaging of the steady NSCH equations in the creeping-flow regime.
The average yields a mean velocity that is consistent with the aggregated forces
acting on the fluid volume. We then construct a potential-flow field that satisfies the
impenetrability condition at domain boundaries and that exhibits the same mean velocity.

\subsection{Intrinsic phase average velocity model}
To derive the multiscale flow model, we conceive of the flow domain~$\Omega\subset\IR^d$ as
the void space of a rigid porous medium.We consider the flow of an incompressible homogeneous Newtonian fluid, subject to a volumetric load $\F:\Omega\to\IR^d$. The velocity~$\U:\Omega\to\IR^d$ and 
pressure $p:\Omega\to\IR$ of such a flow are generally modeled by the Stokes equations:
\begin{equation}
\label{eq:Stokes}
-\mu\Oldnabla^2\U+\nabla{}p=\F\quad\text{in }\Omega
\end{equation}
in combination with the velocity-solenoidality condition $\div\U=0$ and homogeneous boundary
conditions $\U=0$ on the fluid-solid interface. We consider an averaging volume consisting of 
a ball~$B(\X)$ with center~$\X\in\IR^d$ that is sufficiently large with respect to the pore scale. 
Introducing the {\em intrinsic phase average} $\ave{\psi}$ of any 
variable $\psi$ according to:
\begin{equation}
\label{eq:IAve}
\ave{\psi}(\X)=\frac{1}{\mathrm{meas}(B(\X)\cap\Omega)}\int_{B(\X)\cap\Omega}\psi(\X)\,\dd{V}
\end{equation}
the Slattery-Whitaker averaging procedure~\cite{neuman1977theoretical,whitaker1986flow,slattery1999advanced}
leads to the following intrinsic phase average of~\EQ{Stokes}:
\begin{equation}
\label{eq:SWaverage}
\mu\ave{\U}+\varepsilon^{-1}\KK\cdot\big(\nabla\ave{p}-\ave{\F}\big)=0
\end{equation}
where $\KK$ and $\varepsilon$ denote the permeability tensor and porosity, respectively. 
One may note that~\EQ{SWaverage} represents the intrinsic-phase-average form of Darcy's law.

The intrinsic phase average force density $\ave{\F}$ represents the averaged aggregate of forces
on the liquid volume $B(\X)\cap\Omega$, except for the force exerted by the solid medium
in reaction to the viscous and pressure traction, $-\mu\partial_n\U+p\N$ with $\N=-\Nw$ the unit normal
vector exterior to the fluid domain~$\Omega$, which has been 
accounted for in~\EQ{SWaverage} by the slattery-Whitaker averaging procedure. If~\EQ{SWaverage} serves as
a model for~\EQ{NSCHmom}, then $\ave{\F}$ represents the combined effect of the gravitational force 
on the fluid and the tractions exerted by the solid wall in reaction to variations in the phase variable
along~$\Gammaw$. The traction exerted by the solid on the fluid corresponds 
to $\nabla_{\Gamma}\sigmasf(\phi)$,
where $\nabla_{\Gamma}$ denotes the surface gradient along~$\Gammaw$ and~$\sigmasf$ is the solid-fluid surface tension according to~\EQ{sigmasf}. By virtue of the decomposition 
$\nabla=\nabla_{\Gamma}+\N\,\partial_n$ and the wetting boundary condition~\EQ{wettingBC}, the following sequence of identities holds:
\begin{multline}
\nabla_{\Gamma}\sigmasf(\phi)=\sigmasf'(\phi)\nabla_{\Gamma}\phi
=
\sigmasf'(\phi)\nabla\phi-\sigmasf'(\phi)\,\partial_n\phi\,\N
\\
=
-\tilde{\sigma}\epsilon(\partial_n\phi)\nabla\phi+(\tilde{\sigma}\epsilon)^{-1}
(\sigmasf'(\phi))^2\N
\end{multline}
The Slattery-Whitaker averaging procedure relies on the assumption that the fluid-solid contact area is 
significantly larger than the area of the permeable boundary, i.e. 
$\meas(\Omega\cap\partial{}B(\X))\ll\meas(\partial\Omega\cap{}B(\X))$. 
Without further restrictions, we can therefore assume that the following identities hold:
\begin{multline}
\int_{\Gammaw\cap{}B(\X)}\nabla_{\Gamma}\sigmasf(\phi)\,\dd{}S
=
\int_{\partial(\Omega\cap{}B(\X))}
\big(-\tilde{\sigma}\epsilon(\partial_n\phi)\nabla\phi+(\tilde{\sigma}\epsilon)^{-1}
(\sigmasf'(\phi))^2\N\big)\,\dd{}S
\\
=
-\int_{\Omega\cap{}B(\X)}
\div\big(\tilde{\sigma}\epsilon\nabla\phi\otimes\nabla\phi\big)\,\dd{}V
+
\int_{\Gammaw\cap{}B(x)}
(\tilde{\sigma}\epsilon)^{-1}(\sigmasf'(\phi))^2\N\big)\,\dd{}S
\end{multline}
Note that the tensor in the first integral in the ultimate expression coincides with the 
capillary tensor. Hence, we obtain the following expression for the intrinsic phase average
of the aggregated force due to gravity and capillarity:
\begin{equation}
\ave{\F}=\ave{\rho(\phi)}\G-\ave{\div\zeta}+\frac{1}{\meas(B(\X)\cap\Omega)}\int_{\Gammaw\cap{}B(x)}
(\tilde{\sigma}\epsilon)^{-1}(\sigmasf'(\phi))^2\N\,\dd{}S
\end{equation}
Assuming isotropic permeability according to $\KK=\kappa\II$, we then obtain the following 
model for the intrinsic-phase-average velocity:
\begin{multline}
\label{eq:Uave}
\ave{\U}=\frac{\kappa}{\varepsilon\mu}\bigg(-\nabla\ave{p}-\ave{\div\zeta}+\ave{\rho(\phi)}\G
\\
+\frac{1}{\meas(B(\X)\cap\Omega)}\int_{\Gammaw\cap{}B(x)}
(\tilde{\sigma}\epsilon)^{-1}(\sigmasf'(\phi))^2\N\,\dd{}S\bigg)
\end{multline}
It is to be noted that to obtain a macroscopic model of capillary flows in porous media, the surface
term corresponding to the integral on $\Gammaw\cap{}B(\X)$ must be replaced by an appropriate
approximation in intrinsic-phase-average form.

Equation~\EQ{Uave} allows several interpretations, corresponding to various gradations of refinement 
of the model. It can be conceived of as a relation for the global uniform average velocity 
on~$\Omega$, or as a local average velocity field. The proportionality constant $(\varepsilon\mu)^{-1}\kappa$ can be regarded as being uniform in space and time, it can be uniform in 
space but time dependent, e.g. depending on the average saturation $\ave{\phi}$, or it can be 
non-uniform in space and time. In the sequel, we adopt the interpretation of the average velocity in~\EQ{Uave} as a spatially uniform constant vector with a time-dependent constant of proportionality that depends on the mean viscosity.

\subsection{Potential-flow transport velocity}
To determine a velocity field in accordance with~\EQ{Uave} that complies with the impenetrability
condition at the fluid-solid interface, we construct a suitable potential-flow field. The flow 
velocity is given by the gradient of a harmonic potential, $\theta$. The impenetrability
condition at the fluid-solid interface translates into homogeneous Neumann conditions at~$\Gammaw$.
Hence, the potential satisfies:
\begin{subequations}
\label{eq:potential1}
\begin{alignat}{2}
-\Oldnabla^2\theta&=0&\quad&\text{in }\Omega
\label{eq:potential1a}
\\
\partial_n\theta&=0&\quad&\text{on }\Gammaw
\label{eq:potential1b}
\end{alignat}
In addition to~\EQ{potential1}, the potential-flow velocity must be compatible with the average velocity in~\EQ{Uave}, i.e.
\begin{equation}
\label{eq:potential2}
\alpha\cdot\mean{\nabla\theta}=\alpha\cdot\ave{\U}\qquad\forall\alpha\in\IR^d
\end{equation}
\end{subequations}
where for any function $\psi$, $\mean{\psi}$ denotes its average on~$\Omega$ according to
\begin{equation}
\label{eq:mean}
\mean{\psi}=\frac{1}{\mathrm{meas}(\Omega)}\int_{\Omega}\psi\,\dd{}V
\end{equation}
Let us note that in this case it holds that $\ave{\div\zeta}=\mean{\div\zeta}$ and $\ave{\rho(\phi)}=\mean{\rho(\phi)}$, but we retain the distinct notations to indicate that the intrinsic phase average 
and the average on~$\Omega$ are generally different.

The conditions in~\EQ{potential1} do not generally provide a unique definition of~$\theta$. We determine
a suitable solution to~\EQ{potential1} by means of a constrained minimization problem. To provide an ambiance for the minimization problem, we denote by~$H^1(\Omega)$ the usual Sobolev space of square-integrable functions from $\Omega$ into~$\IR$ with square-integrable weak derivatives, by $H^1_*(\Omega)$ the subspace
composed of functions that vanish on average, and by $\smash[tb]{H^1_{0,\Gammaio}(\Omega)}$ 
the subspace of functions that vanish on $\Gammaio:=\partial\Omega\setminus\Gammaw$. It is to be noted 
that~$\Gammaio$ generally corresponds to the in-/outflow part of the boundary.
We consider the following solution to~\EQ{potential1}:
\begin{multline}
\label{eq:thetamin}
\theta=\argmin_{w\in{}H^1_{\star}(\Omega)}
\bigg\{\int_{\Omega}\tfrac{1}{2}|\nabla{}w|\,\dd{}V:\int_{\Omega}\nabla{}w\cdot\nabla\lambda\,\dd{}V=0\;\;\forall\lambda\in{}H^1_{0,\Gammaio}(\Omega),
\\
\alpha\cdot\mean{\nabla{w}}=\alpha\cdot\ave{\U}\;\;\forall\alpha\in\IR^d\bigg\}
\end{multline}
The first constraint in~\EQ{thetamin} corresponds to a weak formulation of~\EQ{potential1}. The second constraint imposes~\EQ{potential2}.

The solution of~\EQ{thetamin} corresponds to a stationary point of the Lagrangian:
\begin{equation}
\label{eq:Lagrangian}
\mathcal{L}(\theta,\lambda,\alpha)
=
\int_{\Omega}\tfrac{1}{2}|\nabla{}\theta|\,\dd{}V
+
\int_{\Omega}\nabla{}\theta\cdot\nabla\lambda\,\dd{}V
+
\alpha\cdot\big(\mean{\nabla{\theta}}-\ave{\U}\big)
\end{equation}
A weak formulation of stationarity of $\mathcal{L}(\theta,\lambda,\alpha)$ in~\EQ{Lagrangian} is provided by:
\begin{multline}
\label{eq:WeakFormLambda}
(\theta,\lambda,\alpha)\in{}H^1_*(\Omega)\times{}H^1_{0,\Gammaio}(\Omega)\times\IR^d:
\\
\int_{\Omega}\nabla(\theta+\lambda+\alpha\cdot\X)\cdot\nabla\delta\theta\,\dd{}V
+
\int_{\Omega}\nabla{}\theta\cdot\nabla\delta\lambda\,\dd{}V
+
\delta\alpha\cdot\mean{\nabla\theta}=\delta\alpha\cdot\ave{\U}
\\
\forall(\delta\theta,\delta\lambda,\delta\alpha)
\in{}H^1_*(\Omega)\times{}H^1_{0,\Gammaio}(\Omega)\times\IR^d
\end{multline}
From~\EQ{WeakFormLambda} we can extract a separate sub-problem for $\lambda$ that facilitates the determination of $\theta$. To this end, we note that for $\delta\lambda=0$ and $\delta\alpha=0$, Equation~\EQ{WeakFormLambda}
implies that $\theta+\lambda+\alpha\cdot\X=\mathrm{const}$. For $\delta\theta=0$ and~$\delta\alpha=0$, Equation~\EQ{WeakFormLambda} implies that $\theta$ is harmonic and satisfies homogeneous Neumann conditions on~$\Gammaw$, in accordance with~\EQ{potential1a} and~\EQ{potential1b}. However, it then follows from $\theta+\lambda+\alpha\cdot\X=\mathrm{const}$ that $\lambda$ is harmonic and $\partial_n\lambda=-\alpha\cdot\N$ on~$\Gammaw$. Setting $\delta\theta=0$ and $\delta\lambda=0$ in~\EQ{WeakFormLambda}, we can infer 
that $\alpha=-\mean{\nabla\lambda}-\ave{\U}$. 
Collecting these results, we obtain the following separate specification for~$\lambda$:
\begin{multline}
\label{eq:lambdarel}
\lambda\in{}H^1_{0,\Gammaio}:\\
\int_{\Omega}\nabla\lambda\cdot\nabla{}\eta\,\dd{V}-\int_{\Gammaw}\eta\mean{\nabla\lambda}\cdot\N\,\dd{}S=
\int_{\Gammaw}\eta\ave{\U}\cdot\N\,\dd{}S
\quad\forall{}\eta\in{}H^1_{0,\Gammaio}
\end{multline}
Once $\lambda$ has been determined from~\EQ{lambdarel}, the gradient of $\theta$ can be formed 
as $\nabla\theta=\mean{\nabla\lambda}+\ave{\U}-\nabla\lambda$.

  \subsection{Non-dimensionalization and weak formulation of the aggregated problem}

  To avoid repetitious reformulation of the equations, we introduce dimensionless groups before
  presenting the complete weak formulation of the aggregated multiscale diffuse-interface model.
  Let $L$ denote a characteristic length scale of the problem under consideration, typically associated with the geometry of the domain. We define the dimensionless domain $\hat{\Omega}=L^{-1}\Omega$. In addition, we define a velocity scale as $U=\kappa\tilde{\sigma}/(\varepsilon\mu{}L^2)$; 
  cf.~\EQ{Uave}. 
  The density and viscosity are non-dimensionalized with respect to the properties of the liquid. Accordingly, we introduce the density ratio $\smash[tb]{\nu_{\rho}=\rhoa/\rhol}$ and viscosity 
  ratio $\smash[tb]{\nu_{\mu}=\mua/\mul}$, and the dimensionless mixture relations:
  \begin{equation}
  \label{eq:rhohatmuhat}
  \hat{\rho}(\phi)=\tfrac{1}{2}(1+\phi)+\tfrac{1}{2}(1-\phi)\nu_{\rho}\qquad
  \hat{\mu}(\phi)=\tfrac{1}{2}(1+\phi)+\tfrac{1}{2}(1-\phi)\nu_{\mu}
  \end{equation}
  The non-dimensionalization procedure conveys the following three dimensionless groups:
    \begin{equation}
    \label{eq:CnBoPe}
      \Cn = \frac{\epsilon}{L} \qquad
      \Bo = \frac{\rhol g L^2}{\tilde{\sigma}} \qquad
      \Pe = \frac{U L^2}{\tilde{\sigma} \gamma} 
    \end{equation}
    For convenience, we will omit the hat symbol that identifies non-dimensional objects in the further exposition.
    
We assume that at the in-/outflow boundary $\Gammaio$ the composition of 
the binary-fluid is described by exogenous data~$\beta$, i.e. $\phi$ satisfies the Dirichlet condition $\phi=\beta$ 
on~$\Gammaio$. Denoting by $\ell_{\beta}\in{}H^1(\Omega)$ a lifting of the Dirichlet data, viz. an arbitrary function in~$H^1(\Omega)$ such that the trace of~$\ell_{\beta}$ on~$\Gammaio$ coincides with~$\beta$,   
the aggregated multiscale diffuse-interface binary-fluid flow model can be condensed into the following weak formulation:
    \begin{multline}  
    \label{eq:WeakForm} 
    (\phi,G,\lambda)\in\big(\ell_{\beta}+H^1_{0,\Gammaio}(\Omega)\big)
\times
H^1(\Omega)
\times
H^1_{0,\Gammaio}(\Omega)
    \\
\int_{\Omega}\chi\,\partial_t\phi\,\dd{}V
+
\int_{\Omega}\chi\,\big(\mean{\nabla\lambda}+\ave{\U}-\nabla\lambda\big)\cdot\nabla\phi\,\dd{}V
+
\frac{1}{\Pe}
\int_{\Omega}\nabla\chi\cdot\nabla{}G\,\dd{}V
\\
+
\int_{\Omega}vG\,\dd{}V
-
\Cn\int_{\Omega}\nabla{}v\cdot\nabla{}\phi\,\dd{}V
-
\frac{1}{\Cn}\int_{\Omega}v\Psi'(\phi)\,\dd{}V
+
\int_{\Gammaw}v\Sigmasf'(\phi)\,\dd{}S
\\
+\int_{\Omega}\nabla\lambda\cdot\nabla{}\eta\,\dd{V}-\int_{\Gammaw}\eta\big(\mean{\nabla\lambda}+\ave{\U}\big)\cdot\N\,\dd{}S
=0
\\
\quad\forall{}(\chi,v,\eta)
\in{}
H^1_{0,\Gammaio}(\Omega)
\times
H^1(\Omega)
\times
H^1_{0,\Gammaio}(\Omega)
    \end{multline}
with $\Sigmasf'=\sigmasf'/\tilde{\sigma}$ and
%
%
\begin{equation}
\label{eq:WeakUave}
\begin{aligned}
\ave{\U}&=-\Cn\big\langle{}\div(\nabla\phi\otimes\nabla\phi)\big\rangle^{\star}
-
\Bo\big\langle\rho(\phi)\big\rangle^{\star}\J
\\
&\quad+
\frac{1}{\Cn}
\frac{1}{\meas(\Omega)}\int_{\Gammaw}\!\!\big(\Sigmasf'(\phi)\big)^2\N\,\dd{}S
\end{aligned}
\end{equation}
Note that $\phi,G$ and $\lambda$ are in fact time-dependent functions, and that the weak formulation~\EQ{WeakForm} is to be understood as a relation that holds for all instances in the considered time interval, $t\in(0,\Tf)$. In addition, $\phi$ must comply with suitable initial conditions, according 
to~$\phi|_{t=0}=\phi_0$ for certain exogenous initial data $\phi_0\in{}H^1(\Omega)$. 

In accordance with our definition of the average velocity $\ave{\U}$ as a uniform vector on~$\Omega$, we can conceive of~$\nabla\ave{p}$ in~\EQ{WeakForm} as a uniform force emanating from a pressure difference that acts on the binary-fluid volume. We will restrict our considerations to $\nabla\ave{p}=0$, which is a meaningful regime for capillarity-driven flows.

\section{Numerical Experiments}  
\label{sec:NumExp}

\subsection{Discretization}
To assess the properties of the presented multiscale diffuse-interface binary-fluid model
for capillarity-driven flows, we consider numerical approximations of~\EQ{WeakForm}. To this end,
we replace the ambient spaces of the weak formulation~\EQ{WeakForm} by finite-element approximation
spaces. In addition, the time-derivative~$\partial_t\phi$ is replaced by a finite-difference approximation.

To enhance the stability of the time-discretization scheme, we apply a splitting 
$\Psi=\Psic-\Psie$ of the double-well potential 
in a convex (contractive) and a concave (expansive) part in accordance with the so-called 
{\em Eyre scheme\/}~\cite{Eyre:1997sp,Elliott:1993sf}:
\begin{equation}
\Psic(\phi)=\phi^2,\qquad
\Psie(\phi)=-\tfrac{1}{4}\phi^4+\tfrac{3}{2}\phi^2-\tfrac{1}{4}
\end{equation}
Let $\{t_n\}_{n=0}^N$ denote a partition of the time interval
$[0,\Tf]$ by a strictly increasing sequence of time instances and let 
$\smash[tb]{V^h_0,W{}^h_0\subset{}H^1_{0,\Gammaio}(\Omega)}$ and~$V^h\subset{}H^1(\Omega)$
denote $H^1$-conforming finite-element approximation spaces, parametrized by the mesh width~$h$
of an underlying finite-element mesh. Let $\smash[tb]{\ell^h_{\beta}}$ denote a suitable finite-element
lifting of the boundary data for the phase variable. 
We consider the following discrete approximation of~\EQ{WeakForm}:
    \begin{multline}  
    \label{eq:WeakForm_h} 
    (\phi^h_n,G^h_n,\lambda^h_n)\in\big(\ell^h_{\beta}+V^h_0\big)
\times
V^h
\times
W^h_0
    \\
\int_{\Omega}\chi^h\,\frac{\phi^h_n}{\Delta{}t_n}\,\dd{}V
+
\int_{\Omega}\chi^h\,\big(\mean{\nabla\lambda^h_n}+\ave{\U}_{n-1})-\nabla\lambda^h_n\big)\cdot\nabla\phi^h_{n}\,\dd{}V
+
\\
\frac{1}{\Pe}
\int_{\Omega}\nabla\chi^h\cdot\nabla{}G^h_n\,\dd{}V
+
\int_{\Omega}v^hG^h_n\,\dd{}V
-
\Cn\int_{\Omega}\nabla{}v^h\cdot\nabla{}\phi^h_n\,\dd{}V
\\
-
\frac{1}{\Cn}\int_{\Omega}v^h\Psic'(\phi^h_n)\,\dd{}V
+
\int_{\Omega}\nabla\lambda^h_n\cdot\nabla{}\eta^h\,\dd{V}-\int_{\Gammaw}\eta^h\mean{\nabla\lambda^h_n}\cdot\N\,\dd{}S
\\
=
\int_{\Omega}\chi{}^h\,\frac{\phi^h_{n-1}}{\Delta{}t_n}\,\dd{}V
-
\frac{1}{\Cn}\int_{\Omega}v^h\Psie'(\phi^h_{n-1})\,\dd{}V
\\
\qquad\qquad+
\int_{\Gammaw}\eta^h\ave{\U}_{n-1}\cdot\N\,\dd{}S
-
\int_{\Gammaw}v^h\Sigmasf'(\phi^h_{n-1})\,\dd{}S
\\
\quad\forall{}(\chi^h,v^h,\eta^h)
\in{}
V^h_0
\times
V^h
\times
W^h_0
    \end{multline}
for $n=1,2,\ldots,N$, where $\Delta{}t_n=t_n-t_{n-1}$ denotes the time step. The initial conditions
are incorporated in~\EQ{WeakForm_h} by identifying $\smash[tb]{\phi^h_0:=\phi_0}$ with the initial data. The
average velocity in~\EQ{WeakForm_h} is defined as
\begin{equation}
\label{eq:WeakUave_h}
\begin{aligned}
\ave{\U}_n=&-\Cn\big\langle{}\div(\nabla\phi_n^h\otimes\nabla\phi_n^h)\big\rangle^{\star}
-
\Bo\big\langle\rho(\phi_n^h)\big\rangle^{\star}\J
\\
&+
\frac{1}{\Cn}
\frac{1}{\meas(\Omega)}\int_{\Gammaw}\!\!\big(\Sigmasf'(\phi^h_n)\big)^2\N\,\dd{}S
\end{aligned}
\end{equation}
cf. Equation~\EQ{WeakUave}. 

The discrete approximation in~\EQ{WeakForm_h} treats 
the expansive part of the double-well potential, the fluid-solid surface tension, 
and the average velocity in an explicit manner. By virtue of the explicit treatment
of the average velocity, the coupling between the equations for $\phi^h_n,G^h_n$ on the one
hand and $\lambda^h_n$ on the other hand is only uni-directional. In particular, $\phi^h_n$ and $G^h_n$
depend on~$\lambda^h_n$, but $\lambda^h_n$ depends only on~$\phi^h_{n-1}$ and $G^h_{n-1}$. Hence,
the system~\EQ{WeakForm_h} can be solved in a partitioned manner. Moreover, one can infer that
the left member of~\EQ{WeakForm_h} corresponds to a bilinear form and, hence, only linear problems
have to be solved within each time step.

\subsection{Flow between two parallel plates}
\label{sec:TC1}
We first consider the application of~\EQ{WeakForm_h} to the generic test case of
flow between two parallel plates. Let us note that in this case the multiscale model is
considered out-of-context, because there is no distinction between the micro scale and the
macro scale. We set $\Omega:=(0,1)\times(0,5)$ and $\Gammaw=\{0,1\}\times(0,5)$.
The separation between the plates has been selected as reference length. 
Gravitational acceleration acts in the direction $-\J=-(0,1)$. We cover~$\Omega$ by a uniform
finite-element mesh $\TT^{\HH}$ with uniform isotropic mesh width $\smash[tb]{h=2^{-6}}$. The finite-element approximation spaces are composed of $C^1$\nobreakdash-continuous piece-wise quadratic polynomials subordinate to the 
mesh~$\TT^{\HH}$ according to
\begin{equation}
V^h=\big\{u\in{}C^1(\Omega):u|_{e}\in{}P^2(e),\;\forall{}e\in\TT^{\HH}\big\}
\end{equation}
where $P^2(e)$ represents the class of bi-variate quadratic polynomials on element~$e$,
and $V^h_0=V^h\cap{}H^1_{0,\Gammaio}(\Omega)$ and~$W^h_0=V^h_0$. The time step is set uniformly to
$\smash[tb]{\Delta{}t=10^{-2}}$.

We regard 4 distinct settings of the dimensionless parameters, as reported in Table~\TAB{TC12}.
The Peclet number, Cahn number and ambient-liquid density ratio and viscosity ratio are identical 
for all test cases.
In the first two cases, the equilibrium contact angles on the left and right wall are identical,
viz. $\thetae^0=\thetae^1=60^{\circ}$, and distinct Bond numbers are considered.
For test cases~C and~D the Bond numbers are identical, but the left and right wall exhibit
distinct equilibrium contact angles. For test case D, in particular, the equilibrium angle at the
left wall exceeds $90^{\circ}$, which is characteristic for a {\em hydrophobic\/} wall.
\begin{table}
\begin{center}
\begin{tabular}{|l|c|c|c|c|c|c|}
\cline{2-7}
\multicolumn{1}{c|}{}&\rule[-5pt]{0pt}{16pt}$\Bo$&$\Cn$&$\thetae^0$&$\thetae^1$&$\Pe$&$\nu_{\rho}$
\\
\hline
\hline
\rule[-5pt]{0pt}{16pt}
case 1A & 0.436 & 0.05 & \multicolumn{2}{c|}{$60^{\circ}$} & $10^3$ & $10^{-3}$ 
\\
\hline
\rule[-5pt]{0pt}{16pt}
case 1B & 0.641 & 0.05 & \multicolumn{2}{c|}{$60^{\circ}$} & $10^3$ & $10^{-3}$ 
\\
\hline
\rule[-5pt]{0pt}{16pt}
case 1C & 0.436 & 0.05 & $60^{\circ}$ & $30^{\circ}$ & $10^3$ & $10^{-3}$ 
\\
\hline
\rule[-5pt]{0pt}{16pt}
case 1D & 0.436 & 0.05 &  $110^{\circ}$ & $30^{\circ}$  & $10^3$ & $10^{-3}$ 
\\
\hline
\hline
\rule[-5pt]{0pt}{16pt}
case 2 & 1.748 & 0.05 & \multicolumn{2}{c|}{$60^{\circ}$} & $10^0$ & $10^{-3}$ 
\\
\hline
\end{tabular}
\caption{Overview of dimensionless parameters for test cases 1A--1D (Section~\SEC{TC1}) and test case 2 (Section~\SEC{TC2}).
\label{tab:TC12}}
\end{center}
\end{table}

%
%

\begin{figure}
\begin{center}
\includegraphics[width=0.8\textwidth]{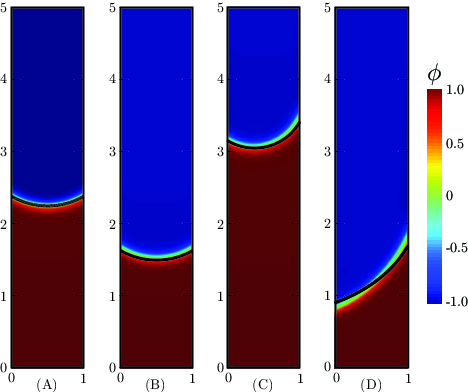}
\caption{Equilibrium phase distribution for test cases 1A--1D (see Table~\TAB{TC12}) and corresponding
sharp-interface reference meniscus shape ({\em solid line\/}). }
\label{fig:TC1}
\end{center}
\end{figure}

Figure~\FIG{TC1} plots the equilibrium phase distributions for test cases~1A--1D. The solid black line
in each panel corresponds to the analytical solution of the meniscus profile according to a sharp-interface model; see~\SEC{MenShape}. The equilibrium height is related to the contact angle, the 
Bond number according to~\EQ{CnBoPe}, and the density ratio by
\begin{equation}
\he=\frac{4\sqrt{2}\,\cos(\thetae)}{3(1-\nu_{\rho})\,\Bo}
\end{equation}
The results in Figure~\FIG{TC1} convey that the equilibrium height and meniscus shape obtained with the multiscale model yield an excellent agreement with the corresponding sharp-interface reference results.
Only for the partly hydrophobic case, viz. case 1D, there is a discernible deviation between the reference result and the result of the multiscale model. We conjecture that this deviation is related to the
relatively wide diffuse interface in combination with the small equilibrium contact angle. Let us note that
the macroscopic (average) meniscus height is very well predicted in all cases.  
  
\subsection{Capillarity-driven flow in a porous medium}  
\label{sec:TC2}
\begin{figure}
\begin{center}
\includegraphics[width=\textwidth]{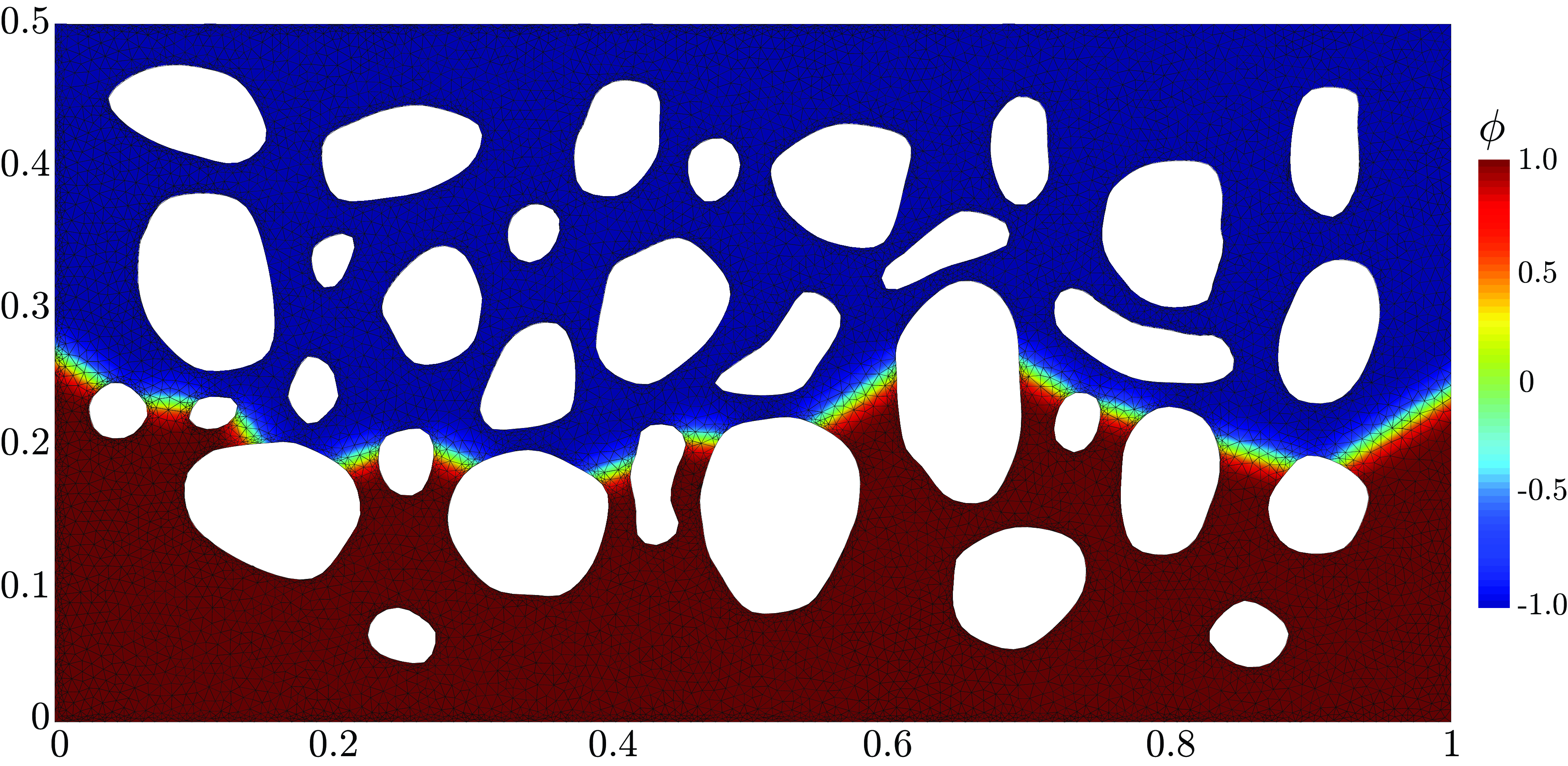}
\caption{Domain with inclusions for test case 2 with finite-element mesh and equilibrium phase field.}
\label{fig:TC2}
\end{center}
\end{figure}

In this section we consider numerical approximations~\EQ{WeakForm_h} for a non-generic geometry, viz. 
a domain with multiple inclusions; see Figure~\FIG{TC2}. The domain is
$\Omega=((0,1)\times(0,1/2))\setminus\mathcal{I}$, where~$\mathcal{I}$ corresponds to a set of
inclusions. We impose Dirichlet boundary conditions on $\Gammaio=(0,1)\times\{0,1/2\}$ 
with data corresponding to the unmixed phases, $\phi=1$ at $x_2=0$ and $\phi=-1$ at~$x_2=1/2$.
The complementary part of the boundary, $\Gammaw=\partial\Omega\setminus\Gammaio$, is provided with the preferential-wetting boundary condition~\EQ{wettingBC}. 
The initial condition is set according to a horizontal meniscus at~$x_2=1/10$:
\begin{equation}
\label{eq:phi0}
\phi_0(\X)=\phie\Big(\frac{-\J\cdot\X+1/10}{\sqrt{2}\epsilon}\Big)
\end{equation}
with $\phie$ the equilibrium profile according to~\EQ{EqProfile}. The dimensionless parameters 
for this test cases are reported in Table~\TAB{TC12}.

We cover~$\Omega$ with an unstructured triangular mesh~$\mathcal{T}^h$ 
as depicted in Figure~\FIG{TC2}. The finite-element
approximation spaces are composed of standard $C^0$\nobreakdash-continuous piece-wise 
linear polynomials subordinate to the mesh~$\mathcal{T}^h$. The time step is set uniformly 
to~$\Delta{}t=5\times{}10^{-2}$.

\begin{figure}
\begin{center}
\includegraphics[width=0.6\textwidth]{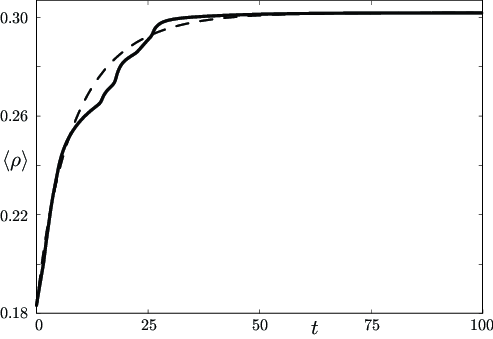}
\caption{Evolution of the average density $\mean{\rho}$ versus time $t$ for test case 2 $(\mbox{---})$ and reference average-density evolution $\rho_{\textsc{e}}(\Delta(t/\Lambda)+1)$ with~$\Delta$ according to~\EQ{Delta} $(--)$ .}
\label{fig:TC2Dynamics}
\end{center}
\end{figure}

Figure~\FIG{TC2Dynamics} displays the evolution of the average density:
\begin{equation}
\mean{\rho}=\frac{1}{\meas{(\Omega)}}\int_{\Omega}\tfrac{1}{2}(1+\phi)+\tfrac{1}{2}(1-\phi)\nu_{\rho}\,\dd{}V
\end{equation}
cf.~\EQ{mean} and~\EQ{rhohatmuhat}. Because at the macroscale level the average density is proportional
to the meniscus height, a reference for the evolution of $\mean{\rho}$ can be determined from~\EQ{Delta}.
The corresponding reference, $\rho_{\textsc{e}}(\Delta(t/\Lambda)+1)$ with $\rho_{\textsc{e}}$ as the
average density in equilibrium, is indicated in Figure~\FIG{TC2Dynamics} by the dashed line. 
The equilibrium average density $\rho_{\textsc{e}}\approx0.302$ has been determined from the 
average density of~$\mean{\rho}$ at $t=100$, which is essentially in equilibrium. The 
corresponding phase field is depicted in Figure~\FIG{TC2}. 
In accordance with the negligible ambient viscosity, 
$\Theta$ in~$\Delta$ according to~\EQ{Delta} has been set to~$0$. The parameter $c\approx-1.33$ in~\EQ{Delta} has been determined from the average density corresponding to the initial condition~\EQ{phi0}, 
i.e.~$c$ satisfies $\rho_{\textsc{e}}W(\exp(c))=\mean{\rho(\phi_0)}$; see~\EQ{Delta}. The time scale
$\Lambda\approx11.5$ has been selected to obtain a visually optimal match between the evolution 
of~$\mean{\rho}$ and the reference. Let us note that the time scale~$\Lambda$ is in fact the only
adjustable parameter, as~$c$ and~$\rho_{\textsc{e}}$ are determined by the initial and equilibrium
properties of~$\mean{\rho}$, respectively. Figure~\FIG{TC2Dynamics} conveys a close correspondence between
the average-density evolution $\mean{\rho}$ and the reference according to $\rho_{\textsc{e}}(\Delta(t/\Lambda)+1)$. The deviations in the interval $t\in(8,40)$ can be attributed
to specific microscale dynamics, related to the shape of the inclusions and, in particular, the evolution of the measure of the contact line, which is in 2D proportional to the cardinality of the set of holes in the meniscus; see Figure~\FIG{TC2}. Because the number of inclusions in the encapsulating 
domain~$(0,1)\times(0,1/2)$ is relatively limited, individual inclusions can bear a significant effect on the macroscopic evolution encoded by~$\mean{\rho}$.

\section{Conclusion and discussion}  
\label{sec:ConcRem}
In this work we presented a multiscale diffuse-interface model for
capillarity-driven flow in porous media. The model is based on averaging
of the equations of motion for a binary fluid with a diffuse interface
according to the Navier-Stokes-Cahn-Hilliard (NSCH) equations. The
resulting model comprises the convective Cahn-Hilliard equations coupled
with a potential-flow transport velocity that complies with the intrinsic 
phase-average velocity of the NSCH equations and the impenetrability condition 
at fluid-solid interfaces. The model is significantly less complex than the 
underlying NSCH equations, as it comprises only the phase-field variable and a
harmonic potential, while it retains the dynamical and equilibrium 
properties of the NSCH equations at a macroscopic level.

The presented model provides a microscopic description of capillarity-driven 
porous-media flow in that it considers the evolution of the meniscus on a 
microscale and explicitly accounts for the geometry of the porous medium. The 
model corresponds to a macroscopic description, on the other hand, in that the 
transport-velocity field is based on conservation of linear momentum in an 
aggregated manner only. In our numerical approximations, we have opted to approximate
the transport-velocity flow potential in the same finite-element space as the 
phase-field and the chemical potential in the Cahn-Hilliard equations. Two
opposite extensions of the model can be envisaged. On the one hand, the transport-velocity
field can be even further reduced, for instance, by approximating it in a coarser
finite-element approximation space. From a macroscopic perspective, the main property 
of the transport velocity is that it is proportional to the intrinsic phase-average
of the volumetric and traction forces. The model for the transport-velocity potential
can be significantly reduced without violating this property. On the other hand,
the transport-velocity model can be extend with a more localized representation of conservation
of linear momentum to obtain a more precise approximation of the microscopic dynamics,
e.g. by requiring that local averages of the transport velocity are proportional to local
averages of the forces, or by requiring that additional moments of the transport velocity are
proportional to the corresponding moments of the volumetric and traction forces.

\section*{Acknowledgment}
This work is supported by NanoNextNL, a micro and nanotechnology consortium of the Government of the Netherlands and 130 partners.

\begin{appendix}

\section{}
\label{sec:MenShape}
\begin{figure}
\begin{center}
\includegraphics[width=0.35\textwidth]{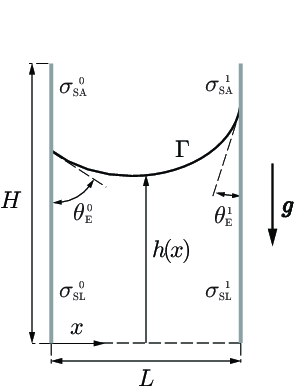}
\caption{Illusration of capillary rise between two walls with dissimilar wetting properties.}
\label{fig:IlluCap}
\end{center}
\end{figure}

To enable a detailed validation of the multiscale diffuse-interface binary-fluid flow model,
we derive here an expression for the equilibrium shape of the meniscus for the general case of a binary-fluid flow between to parallel surfaces with distinct wetting properties, encoded by distinct equilibrium
contact angles. To facilitate the presentation, Figure~\FIG{IlluCap} illustrates the considered problem.
We consider two parallel surfaces, separated by a distance~$L$. We assume that gravity acts vertically downward, parallel to the surfaces. Let $x$ then denote a horizontal coordinate, centered
at the left surface, and let $h:(0,L)\to\IR$ designate the height of the meniscus, i.e. the meniscus
is parametrized according to $\Gamma:=\{(x,y)\in\IR^2:y=h(x),\:0\leq{}x\leq{}L\}$. 
The thermodynamic potential associated with a particular height function $h$ writes~\cite{bullard2009capillary}:
\begin{multline}
\label{eq:Emeniscus}
E(h)=
\sigmasa^0\big(H-h(0)\big)+\sigmasl^0h(0)
+
\sigmasa^1\big(H-h(L)\big)+\sigmasl^1h(L)
\\
+
\int_0^L\sigma\big(1+(Dh(x))^2\big)^{1/2}\dd{}x
+
\frac{1}{2}
\bigg(
\int_0^L\rhol{}gh^2(x)\,\dd{}x
+
\int_0^L\rhoa{}g\big(H-h(x)\big)^2\,\dd{}x\bigg)
%
\end{multline}
where $D(\cdot)$ denotes the derivative of~$(\cdot)$.
The terms with $\sigmasa$ and $\sigmasl$ express the energy contributions from the solid-fluid
interfaces and, in particular, the solid-ambient and solid-liquid contributions. The superscript~0
(resp.~1) indicates contributions from the left (resp. right) surface; see Figure~\FIG{IlluCap}. 
The first integral term 
corresponds to the surface-tension contribution from the meniscus. The penultimate and ultimate
terms are associated with the gravitational-energy contributions of the liquid and ambient fluid. 

We define a non-dimensionalized coordinate $\smash[tb]{\hat{x}=x/L}$ and meniscus 
elevation $\smash[tb]{\hat{h}=h/L}$. 
Introducing the dimensionless energy functional $\smash[tb]{\hat{E}=E/\sigma{}L}$, the liquid-ambient mass
ratio $\nu_{\rho}=\rhoa/\rhol$ and
the Bond number $\BoSharp=\rhol{}gL^2/\sigma$, the expression for the
free energy of the meniscus shape can be recast in non-dimensionalized form as
\begin{multline}
\label{eq:hatE}
\hat{E}(\hat{h})
=
\hat{E}_0
-
\cos(\thetae^0)\,\hat{h}(0)
-
\cos(\thetae^1)\,\hat{h}(1)
+
\int_0^1\big(1+\big(D\hat{h}(\hat{x})\big)^2\big)^{1/2}\dd{}\hat{x}
\\
+
\frac{\BoSharp}{2}
\bigg(
\int_0^1\hat{h}^2(\hat{x})\,\dd{}\hat{x}
+
\int_0^1\nu_{\rho}\big(\hat{H}-\hat{h}(\hat{x})\big)^2\,\dd{}\hat{x}\bigg)
\end{multline}
where $\hat{E}_0:=\sigma^{-1}(\sigmasa^0+\sigmasa^1)\hat{H}$ represents the 
$\hat{h}$\nobreakdash-independent contribution to the energy due to the height of the surfaces.
Let us note that the Bond number based on the sharp-interface fluid-fluid surface tension $\BoSharp$
and the Bond number for the diffuse-interface model~$\Bo$ in~\EQ{CnBoPe} are related by $3\,\Bo=2\sqrt{2}\,\BoSharp$, on account of the relation between~$\tilde{\sigma}$ and~$\sigma$.

The optimality conditions associated with~\EQ{hatE} can be phrased as:
\begin{multline}
\label{eq:DhatE}
-
\cos(\thetae^0)\,{v}(0)
-
\cos(\thetae^1)\,{v}(1)
+
\int_0^1\big(1+\big(D{h}({x})\big)^2\big)^{-1/2}D{h}({x})\,D{v}(x)\,\dd{}{x}
\\
+
\BoSharp
\bigg(
\int_0^1{h}({x}){v}({x})\,\dd{}{x}
-
\int_0^1\nu_{\rho}\,\big({H}-{h}({x})\big){v}({x})\,\dd{}{x}\bigg)
=0
\end{multline}
for all admissible ${v}$, where the hat symbols have been omitted for transparency. 
Integration-by-parts conveys that~\EQ{DhatE} implies the following 2-point boundary-value problem:
\begin{subequations}
\label{eq:hBVP}
\begin{alignat}{2}
-D\bigg(\frac{Dh}{({1+(Dh)^2})^{1/2}}\bigg) + 
\BoSharp(1+\nu_{\rho})h&= \BoSharp\,\nu_{\rho}H
&\qquad&\text{in }(0,1)
\\
Dh&=\frac{-1}{\tan(\thetae^0)}
&\qquad&\text{at }\{0\}
\\
Dh&=\frac{1}{\tan(\thetae^1)}
&\qquad&\text{at }\{1\}
\end{alignat}
\end{subequations}

Boundary-value problem~\EQ{hBVP} can be solved in closed form for certain specific
cases~\cite{de2013capillarity}. To plot the reference meniscus shapes
in Section~\SEC{NumExp}, we construct a finite-element approximation to~\EQ{hBVP} based on~\EQ{DhatE}.
For the solution of this non-linear problem, we apply a Picard iteration procedure.

\end{appendix}

\section*{References}
\bibliographystyle{elsarticle-num} 
\bibliography{CapPaper}

\begin{thebibliography}{10}
\expandafter\ifx\csname url\endcsname\relax
  \def\url#1{\texttt{#1}}\fi
\expandafter\ifx\csname urlprefix\endcsname\relax\def\urlprefix{URL }\fi
\expandafter\ifx\csname href\endcsname\relax
  \def\href#1#2{#2} \def\path#1{#1}\fi

\bibitem{vafai2005handbook}
K.~Vafai, Handbook of porous media, Crc Press, 2005.

\bibitem{coutelieris2012transport}
F.~A. Coutelieris, J.~M. Delgado, Transport processes in porous media, Vol.~20,
  Springer Science \& Business Media, 2012.

\bibitem{bear2013dynamics}
J.~Bear, Dynamics of fluids in porous media, Courier Corporation, 2013.

\bibitem{dybbs1984new}
A.~Dybbs, R.~Edwards, A new look at porous media fluid mechanics - {D}arcy to
  turbulent, in: Fundamentals of transport phenomena in porous media, Springer,
  1984, pp. 199--256.

\bibitem{Darcy:1856yu}
H.~Darcy, Les fontaines publiques de la ville de Dijon, Victor Dalmont, 1856.

\bibitem{neuman1977theoretical}
S.~P. Neuman, Theoretical derivation of {D}arcy's law, Acta Mechanica 25~(3-4)
  (1977) 153--170.

\bibitem{whitaker1986flow}
S.~Whitaker, Flow in porous media i: A theoretical derivation of {D}arcy's law,
  Transport in porous media 1~(1) (1986) 3--25.

\bibitem{papatzacos2002macroscopic}
P.~Papatzacos, Macroscopic two-phase flow in porous media assuming the
  diffuse-interface model at pore level, Transport in porous media 49~(2)
  (2002) 139--174.

\bibitem{Szymkiewicz:2013jk}
A.~Szymkiewicz, Modelling water flow in unsaturated porous media: accounting
  for nonlinear permeability and material heterogeneity, Springer Science \&
  Business Media, 2012.

\bibitem{emmerich2003diffuse}
H.~Emmerich, The diffuse interface approach in materials science: thermodynamic
  concepts and applications of phase-field models, Vol.~73, Springer Science \&
  Business Media, 2003.

\bibitem{anderson1998diffuse}
D.~Anderson, G.~B. McFadden, A.~Wheeler, Diffuse-interface methods in fluid
  mechanics, Annual review of fluid mechanics 30~(1) (1998) 139--165.

\bibitem{lowengrub1998quasi}
J.~Lowengrub, L.~Truskinovsky, Quasi--incompressible {C}ahn--{H}illiard fluids
  and topological transitions, in: Proceedings of the Royal Society of London
  A: Mathematical, Physical and Engineering Sciences, Vol. 454, The Royal
  Society, 1998, pp. 2617--2654.

\bibitem{Korteweg:1901kn}
D.~Korteweg, Sur la forme que prennent les \'equations du mouvement des fluids
  si l'on tient compte des forces capillaires caus\'es par les variations de
  densit\'e, Archives N\'eerlandaises des Science Exactes et Naturelle, Serie
  II 6 (1901) 1--24.

\bibitem{Liu:2013vn}
J.~Liu, H.~Gomez, J.~A. Evans, T.~J. Hughes, C.~M. Landis, Functional entropy
  variables: a new methodology for deriving thermodynamically consistent
  algorithms for complex fluids, with particular reference to the isothermal
  {N}avier--{S}tokes--{K}orteweg equations, Journal of Computational Physics
  248 (2013) 47--86.

\bibitem{Jacqmin1999calculation}
D.~Jacqmin, Calculation of two-phase {N}avier--{S}tokes flows using phase-field
  modeling, Journal of Computational Physics 155~(1) (1999) 96--127.

\bibitem{Yue:2006comp}
P.~Yue, C.~Zhou, J.~J. Feng, A computational study of the coalescence between a
  drop and an interface in {N}ewtonian and viscoelastic fluids, Physics of
  Fluids (1994-present) 18~(10) (2006) 102102.

\bibitem{seppecher1996moving}
P.~Seppecher, Moving contact lines in the {C}ahn-{H}illiard theory,
  International journal of engineering science 34~(9) (1996) 977--992.

\bibitem{jacqmin2000contact}
D.~Jacqmin, Contact-line dynamics of a diffuse fluid interface, Journal of
  Fluid Mechanics 402 (2000) 57--88.

\bibitem{yue2011wall}
P.~Yue, J.~J. Feng, Wall energy relaxation in the {C}ahn--{H}illiard model for
  moving contact lines, Physics of Fluids (1994-present) 23~(1) (2011) 012106.

\bibitem{Brummelen:2015qv}
E.~H. van Brummelen, H.~Shokrpour~Roudbari, G.~J. van Zwieten,
  Elasto-capillarity simulations based on the
  {N}avier-{S}tokes-{C}ahn-{H}illiard equations, ArXiv (2015) 1--8.

\bibitem{huh1971hydrodynamic}
C.~Huh, L.~Scriven, Hydrodynamic model of steady movement of a
  solid-liquid-fluid contact line, Journal of Colloid and Interface Science
  35~(1) (1971) 85--101.

\bibitem{jacqmin2004onset}
D.~Jacqmin, Onset of wetting failure in liquid--liquid systems, Journal of
  Fluid Mechanics 517 (2004) 209--228.

\bibitem{Cahn:1958fk}
J.~W. Cahn, J.~E. Hilliard, Free energy of a nonuniform system. i. interfacial
  free energy, The Journal of chemical physics 28~(2) (1958) 258--267.

\bibitem{Cueto-Felgueroso:2009ud}
L.~Cueto-Felgueroso, R.~Juanes, Stability analysis of a phase-field model of
  gravity-driven unsaturated flow through porous media, Phys. Rev. E 79 (2009)
  036301.

\bibitem{Cueto-Felgueroso:2009dn}
L.~Cueto-Felgueroso, R.~Juanes, A phase field model of unsaturated flow, Water
  Resour. Res. 45 (2009) W10409.

\bibitem{de2013capillarity}
P.-G. De~Gennes, F.~Brochard-Wyart, D.~Qu{\'e}r{\'e}, Capillarity and wetting
  phenomena: drops, bubbles, pearls, waves, Springer Science \& Business Media,
  2013.

\bibitem{Corless:1996kl}
R.~M. Corless, G.~H. Gonnet, D.~E.~G. Hare, D.~J. Jeffrey, D.~E. Knuth, On the
  {L}ambert {W} function, Adv. Comput. Math. 5 (1996) 329--359.

\bibitem{Taylor:1994uk}
J.~E. Taylor, J.~W. Cahn, Linking anisotropic sharp and diffuse surface motion
  laws via gradient flows, Journal of Statistical Physics 77~(1-2) (1994)
  183--197.

\bibitem{Abels:2012vn}
H.~Abels, H.~Garcke, G.~Gr{\"u}n, Thermodynamically consistent, frame
  indifferent diffuse interface models for incompressible two-phase flows with
  different densities, Mathematical Models and Methods in Applied Sciences
  22~(03) (2012) 1150013.

\bibitem{slattery1999advanced}
J.~C. Slattery, Advanced transport phenomena, Cambridge University Press, 1999.

\bibitem{Eyre:1997sp}
D.~J. Eyre, An unconditionally stable one-step scheme for gradient systems,
  Unpublished article.

\bibitem{Elliott:1993sf}
C.~M. Elliott, A.~Stuart, The global dynamics of discrete semilinear parabolic
  equations, SIAM journal on numerical analysis 30~(6) (1993) 1622--1663.

\bibitem{bullard2009capillary}
J.~W. Bullard, E.~J. Garboczi, Capillary rise between planar surfaces, Physical
  Review E 79~(1) (2009) 011604.

\end{thebibliography}

\end{document}